\documentclass[%
 reprint,
 amsmath,amssymb,
 aps,
]{revtex4-2}

\usepackage{graphicx}
\usepackage{dcolumn}
\usepackage{bm}
\usepackage{lineno}
\usepackage{float}
\UseRawInputEncoding

\begin{document}

\preprint{APS/123-QED}
\raggedbottom

\title{Surface acoustic wave generation and detection in quantum paraelectric regime of SrTiO$_3$-based heterostructure}

\author{Dengyu Yang$^{1, 2}$}
\author{Muqing Yu$^{1, 2}$}
\author{Yun-Yi Pai$^{1, 2}$}
\author{Patrick Irvin$^{1, 2}$}
\author{Hyungwoo Lee$^3$}
\author{Kitae Eom$^3$}
\author{Chang-Beom Eom$^3$}
\author{Jeremy Levy$^{1, 2}$}
 \email{jlevy@pitt.edu}
\affiliation{
1. Department of Physics and Astronomy, University of Pittsburgh, Pittsburgh, Pennsylvania 15260, USA\\
}%
\affiliation{
2. Pittsburgh Quantum Institute, Pittsburgh, Pennsylvania 15260, USA\\
}%
\affiliation{
3. Department of Materials Science and Engineering, University of Wisconsin–Madison, Madison, Wisconsin 53706, USA
}

\date{\today}

\begin{abstract} 
Strontium titanate (STO), apart from being a ubiquitous substrate for complex-oxide heterostructures, possesses a multitude of strongly-coupled electronic and mechanical properties. Surface acoustic wave (SAW) generation and detection offers insight into electromechanical couplings that are sensitive to quantum paraelectricity and other structural phase transitions. Propagating SAWs can interact with STO-based electronic nanostructures, in particular LaAlO$_3$/SrTiO$_3$ (LAO/STO). Here we report generation and detection of SAW within LAO/STO heterointerfaces at cryogenic temperatures ($T\ge$~2 K) using superconducting interdigitated transducers (IDTs). The temperature dependence shows an increase in the SAWs quality factor that saturates at $T\approx 8 $ K. The effect of backgate tuning on the SAW resonance frequency shows the possible acoustic coupling with the ferroelastic domain wall evolution. This method of generating SAWs provides a pathway towards dynamic tuning of ferroelastic domain structures, which are expected to influence electronic properties of complex-oxide nanostructures. Devices which incorporate SAWs may in turn help to elucidate the role of ferroelastic domain structures in mediating electronic behavior.
\end{abstract}

\maketitle

\section{\label{sec:level1}Introduction}
Strontium titanate holds a unique place among the growing family of complex-oxide heterostructures and nanostructures \cite{Pai2018-br}. Apart from possessing a wealth of physical phenomena--ferroelectricity \cite{Burke1971-mz, Tikhomirov2002-wk}, ferroelasticity \cite{Buckley1999-lm, Honig2013-uh}, superconductivity \cite{Reyren2007-ho, Caviglia2008-bf,Ueno2008-al}, high spin-to-charge interconversion \cite{Lesne2016-le}, large third-order optical susceptibility \cite{Ma2013-hv} -- STO also exhibits fascinating transport properties at interfaces and within conductive nanostructures \cite{Annadi2018-az, Cheng2015-fv}. These latter properties arise when STO is capped with a thin layer, often but not exclusively LaAlO$_3$, which results in electron doping near the STO interface \cite{Ohtomo2004-ed}. Conductive nanostructures of many types have been created by ``sketching" with a conductive atomic force microscope (c-AFM) tip \cite{Cen2009-ja} or ultra-low-voltage focused electron beam \cite{doi:10.1063/5.0027480}. The properties of these devices are profoundly affected by the intrinsic behavior of STO and are in many aspects not well understood.

One of the least well-understood property interrelationships concerns the coupling between electronic, ferroelectric, and ferroelastic degrees of freedom.
STO is centrosymmetric at room temperature with \textit{ABO$_3$} cubic perovskite structure. Upon cooling below $\sim$105 K \cite{Shirane1969-hp,Fleury1968-qs}, STO undergoes a cubic-to-tetragonal antiferrodistortive (AFD) phase transition \cite{Whangbo2015-sx}. Further cooling to $\sim$35 K \cite{Cowley1964-it} gives rise to an incipient ferroelectric or ``quantum paraelectric" phase transition (QPE) in which the dielectric constant $\varepsilon$ saturates at $\sim$10 K \cite{Muller1979-uj}. At cryogenic temperature ($T < 10$ K), STO shows giant piezoelectricity even larger than the best well-known piezoelectric material such as quartz \cite{Grupp1997-de}. At microscopic scales, the coupling between polar phases in STO and ferroelastic domains \cite{Salje2016-no} is quite strong and can be directly observed using scanning single electron transistor microscopy \cite{Honig2013-uh}. Piezoelectric distortions were found to be the result of reorienting tetragonal domains, whose in-plane and out-of-plane lattice constants differ by $\sim 10^{-3}$.

Surface acoustic waves (SAW), also known as Rayleigh waves \cite{Rayleigh1885}, arise from linear piezoelectric coupling, and propagate parallel to the sample surface with its depth comparable to the SAW wavelength. SAW propagation is sensitive to both mechanical and electrical changes at the sample surface, making it surface-sensitive and useful for radio-frequency (RF) signal processing. A common technique to generate and detect SAW is to apply RF signals to a pair of metallic inter-digitated transducers (IDT). 
However, the complexity and subtlety of the STO structure with multiple phases make SAW generation and detection difficult to achieve. 
With STO, DC fields have been used to break cubic symmetry and generate polarization above 150 K \cite{Uzun2020,Alzuaga2014} via electrostrictive effect. To generate SAW, an extra piezoelectric layer, PZT, was deposited on top of LAO \cite{Uzun2020PZT}, and SAW was observed down to $T=$110 K. Below this temperature, the signal disappeared and SAW generation and detection has not to our knowledge been reported in STO or LAO/STO or at temperatures lower than $T=$105 K. 

In this paper, we demonstrate direct SAW generation and detection on LAO/STO surface at cryogenic temperatures using superconducting IDTs. The linearly-coupled SAW shows an ultra-low phase velocity, indicating softening of STO crystal at low temperature and consistent with earlier reports of large piezoelectric and electrostriction coefficients \cite{Grupp1997-de}. The temperature at which the quality factor of the SAW resonator saturates coincides with the quantum-paraelectric (QPE) transition temperature ($T_\textrm{QPE}$), showing that the quality factor $Q$ is coupled to the dielectric constant and can be used to identify the onset of the quantum paraelectric phase. The resonance frequency can be tuned with a backgate voltage. The tunability with applying the backgate at negative side but not at the positive backgate side coincides with the tuning effect of ferroelastic domain with the backgate showing the coupling between ferroelastic domains and surface phonon. The applied DC bias confirms the electrostrictive effect from STO by showing the quadratic tuning behavior.

\section{\label{sec:level1}Experiment}

\begin{figure*}
\includegraphics{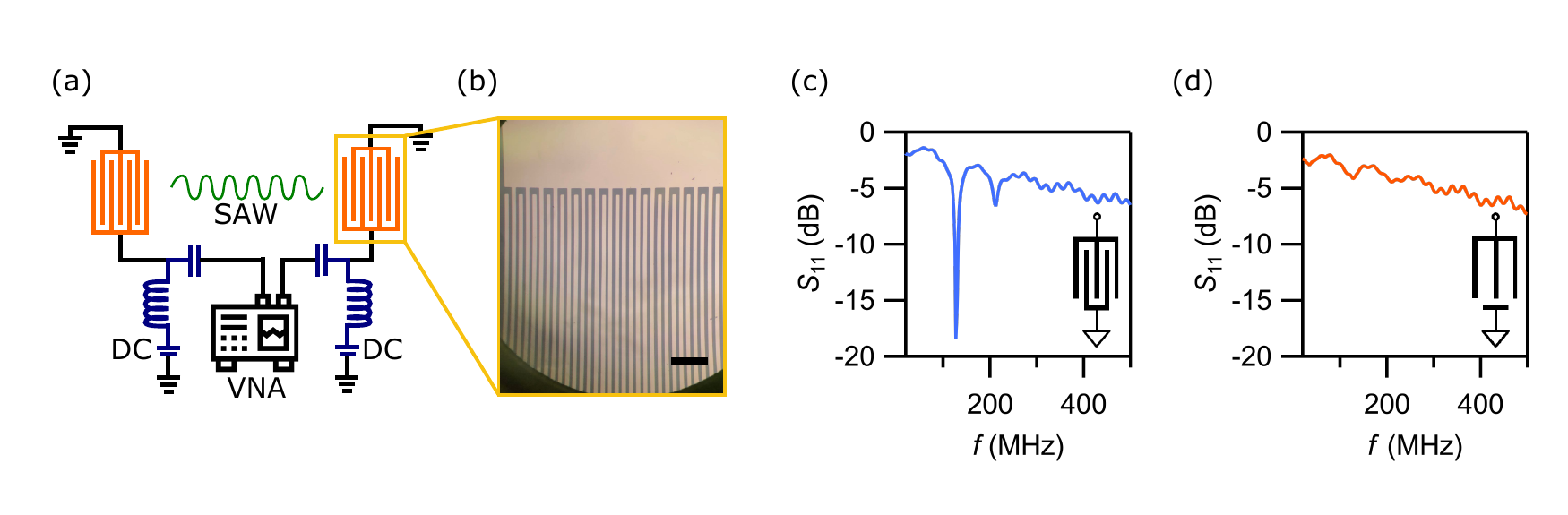}
\caption{Surface Acoustic Wave (SAW) generated and detected by Interdigitated Transducers (IDTs). (a) Schematic diagram of the experiment setup. The orange parts are IDTs. Blue circuits denote the bias tee inserted between vector network analyzer (VNA) port and IDT. (b) Optical image of an IDT patterned with NbTiN. The black scale bar is 20 $\mu \textrm{m}$. (c) $S_{11}$ from an experiment device with normal IDT comb structure in pair. The resonances is observable. (d) Reflection signal $S_{11}$ from a control device without one side of IDT comb structure. There is no resonance observed from this control device.}
\label{fig:epsart}
\end{figure*}

LaAlO$_3$ epitaxial films were grown on TiO$_2$-terminated STO (001) substrates by pulsed laser deposition \cite{Ohtomo2004-ed}. The thickness of LAO is fixed to 3.4 u.c., close to the critical thickness of metal-insulator transition \cite{Thiel2006-ds}. To form a uniform-type single electrode IDT, an 80 nm thick film of NbTiN is deposited on top of the LAO/STO, with IDT fingers oriented along the (010) direction. Superconducting NbTiN is chosen as the IDT material for three principal reasons: to help with impedance matching; to maximize the transmission; and to minimize ohmic losses and heating. A metallization ratio ($m$), defined as the finger width divided by the finger spacing, $m\equiv w/(w+d)$, is fixed such that $m=0.5$ in all devices. SAW-related experiments are carried out in a physical property measurement system (PPMS) at temperatures $T\geq$ 2 K. Each IDT is grounded on one side, and the other side is connected to an input port of a vector network analyzer (VNA) to enable two-port scattering parameter measurements (Fig.~\ref{fig:epsart}(a)). Between the IDT and the VNA, a bias tee is inserted on each side to allow a DC bias to be ($V_\textrm{bias}$) applied between the IDT fingers. SAWs are generated by an IDT, transmitted along the (100) direction, and detected by the second IDT pair. To reduce contributions from bulk acoustic waves, the LAO/STO sample bottom surface is roughened and coated with silver epoxy as a ``soft conductor" \cite{CAMPBELL19899}. The bottom conducting electrode is also used to apply a voltage $V_\textrm{bg}$ from the back of the STO substrate.

Using a $P=-10$ dBm signal applied to the IDT, a clear resonant feature can be seen at 127.5 MHz in the reflection spectrum $S_\textrm{11}$ (Fig.~\ref{fig:epsart}(c)), defined as the center frequency $f_\textrm{c}$. By contrast, in a control device in which one side of the two comb structures in the IDT is missing there is no resonance (Fig.~\ref{fig:epsart} (d)), demonstrating that the resonance feature is a result of the paired comb structures patterned with NbTiN, and not due to bulk acoustic wave transmission or an electrical resonance from the cable or other parts of the instrument. Meanwhile, the SAW phase velocity is obtained from the measured $f_\textrm{c}$ by $v = f_\textrm{c}\lambda$. The wavelength $\lambda$ is determined by the distance between a pair of nearest IDT fingers with the same polarity. Here we have $\lambda = 8~\mu \textrm{m}$, giving a SAW velocity on LAO/STO of $v=1,020$ m/s. 
The IDT comb structure generates SAW by converting the electrical energy to elastic energy, causing a resonance dip in the reflection signal. 

The total quality factor $Q$ is defined as 
\begin{equation}\label{eq:1}
Q\equiv f_\textrm{c}/B,
\end{equation}
where $f_\textrm{c}$ is the center resonance frequency and $B$ is the half-power (-3 dB) bandwidth. The resonance spectrum shows a quality factor $Q = 17.5$, which is consistent with previous reports \cite{Alzuaga2014} on STO-based acoustic resonators without resonance-enhanced structures (e.g., Bragg mirrors). Theoretically the bandwidth $B$ can be determined from the IDT geometry according to Ref. \cite{Aref2016},
\begin{equation}\label{eq:2}
    B\sim0.9f_\textrm{c}/N_\textrm{p},
\end{equation}
where $N_\textrm{p}$ = 16 is the number of comb pairs in the IDT. Here the calculated bandwidth is 7.2 MHz is close to the expected value of 7.3 MHz. The $<2\%$ difference can come from the imperfect edge of IDT geometry related to the mask-less photolithography precision. 

\begin{figure*}
\includegraphics{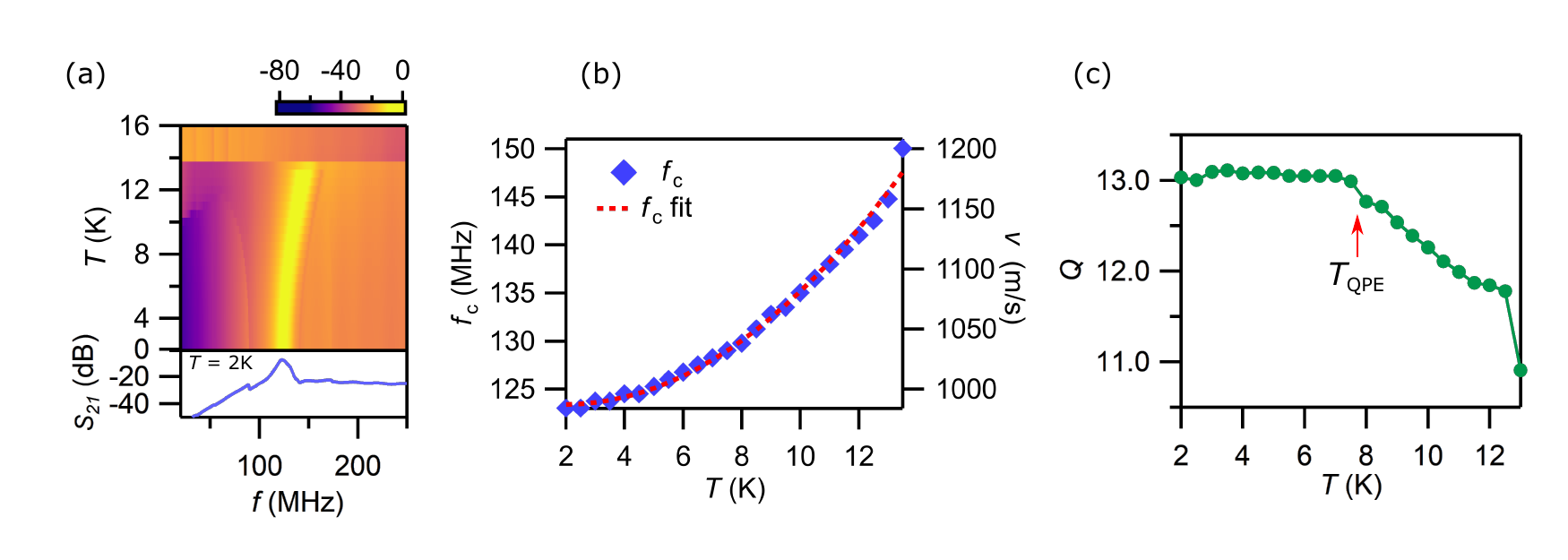}
\caption{Temperature dependence of resonance. (a) The upper intensity plot shows transmission signals with respect to the temperature between 2 K and 16 K. The lower figure is a transmission signal line cut through 2 K temperature showing the resonance peak. (b) Temperature dependence of resonance center frequency and calculated SAW phase velocity (blue diamonds). The red dashed line is a quadratic fit. (c) Quality factor $Q = f_\textrm{c}/B$ plotted with respect to the temperature. The red arrow shows the STO quantum paraelectric saturation temperature.}
\label{fig:Tdep}
\end{figure*}

The transmission spectrum $S_\textrm{21}$ (Fig.~\ref{fig:Tdep} (a)) shows a resonance peak at a frequency $f_\textrm{c}$ that coincides with the reflection dip in $S_\textrm{11}$, supporting the scenario that energy is transmitted efficiently via SAW from the transmitting IDT to the receiving IDT. When we sweep the temperature, the resonance peak disappears sharply at temperatures larger than 13.7 K, corresponding to the temperature $T_\textrm{c}$ above which the NbTiN is no longer superconducting (see Supplementary material). The NbTiN normal resistance 1.57 k$\Omega$ gives an impedance mismatch which causes most of the power to be reflected and dissipated both internally in the IDT and externally into the transmission line leading to a sharp cut-off on the transmission signal.

Notably, the resonance frequency is temperature-dependent. 
A quadratic scaling is observed between the center frequency and temperature, with a lower $f_\textrm{c}$ at a lower temperature.
To understand this scaling, we may consider a Helmholtz free energy of phonons, $F$, description of its equilibrium state,
\begin{equation}
    \label{eq:3}
        F (t, \psi) =  a(t) + b(t)\psi^2 + c(t)\psi^4 + \cdots,
\end{equation}
where $\psi$ is the order parameter, and $t = (T-T_c)/T_c$ is the reduced temperature.
When we only consider the equilibrium states, we obtain \[b(t)\psi+2c(t)\psi^3 = 0\]
\[|\psi| \approx (b_1/2c_0)|t|^{1/2} . \]
The asymptotic expression for $F$ becomes:
\begin{equation} \label{eq:4}
    F (t, \psi)  \approx a_0 - \frac{b_1^2}{4c_0}t^2 + \cdots
\end{equation}
Therefore, the free energy is expected to scale quadratically with temperature. The resonance frequency $f_c$, depending linearly on $F$, scales quadratically with temperature (Fig.~\ref{fig:Tdep} (b)). 

With a constant IDT geometry, such that the wavelength $\lambda$ is kept fixed, a smaller $f_\textrm{c}$ corresponds to a lower SAW phase velocity $v$. We observe that lowering the temperature reduces the SAW velocity. This trend contrasts with results on most piezoelectric materials such as PZT, in which SAW have been reported to increase as temperature is lowered, because of the decreasing thermal fluctuations and increasing stiffness of the sample \cite{doi:10.1063/1.1636530}. 

Both the piezoelectric coefficient ($d_\textrm{311}$) and the electrostriction coefficient ($R_\textrm{311}$) of STO increase significantly with decreasing temperature, especially for $T<$ 10 K \cite{Grupp1997-de}. This finding implies both a softer crystal and a more efficient electro-mechanical energy conversion at low temperature. 
The quality factor, plotted versus temperature in Figure~\ref{fig:Tdep} (c) first increases as temperature is decreased, and then saturates at $T \approx 8$ K. The saturation temperature coincides with the STO quantum paraelectric phase transition ($T_\textrm{QPE}$) where the dielectric permittivity $\varepsilon$ saturates, described by Barrett's formula \cite{Muller1979-uj}. This correspondence indicates that SAW is sensitive to the quantum paraelectric phase transition and its $Q$ factor is related to the $\varepsilon$ variance. When $T > T_c$, Q drops quickly to near zero due to the increasing resistance $R$ for the IDT.

\begin{figure*}
\includegraphics{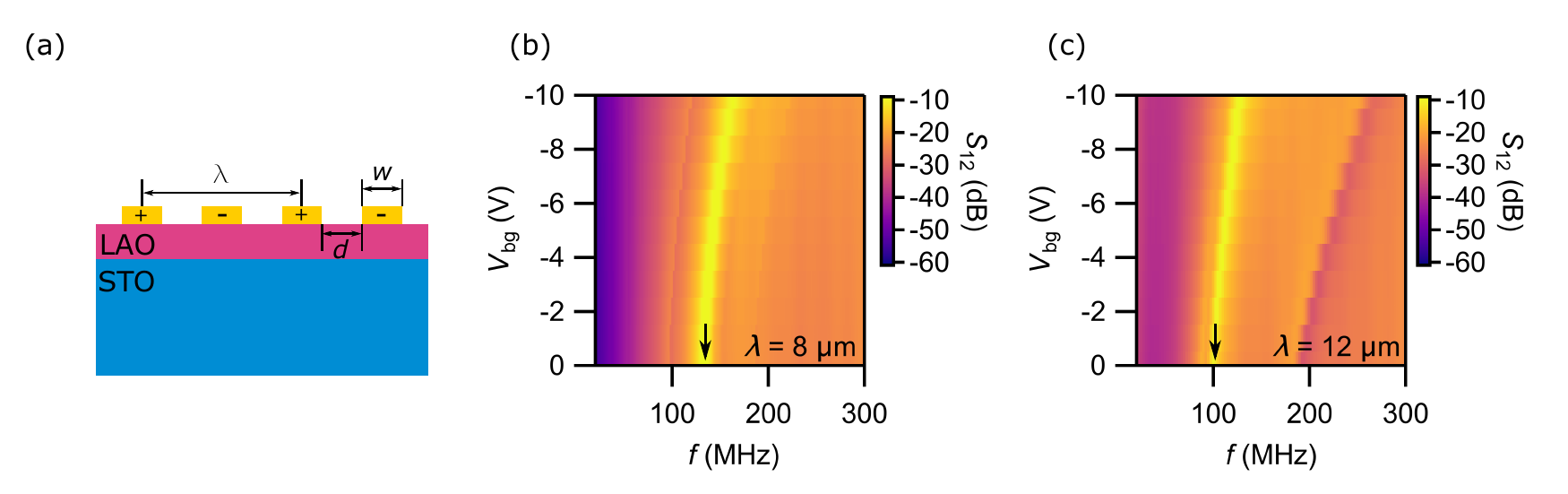}
\caption{Resonance frequency shift due to different wavelengths and negative backgate voltages. (a) Schematic diagram of NbTiN IDT geometry, where $w$ is the finger width and $d$ is the gap distance. Wavelength $\lambda$ is determined by the center distance between two nearest same polarity fingers. (b) Transmission spectrum of Device ``A" ($\lambda = 8~\mu \textrm{m}$) as a function of $V_\textrm{bg}$. Black arrow denotes the resonance frequency. (c) Transmission spectrum of Device ``B" ($\lambda = 12~\mu \textrm{m}$) as a function of $V_\textrm{bg}$. Black arrow denotes the resonance frequency. All data taken at $T = 2~\textrm{K}$}
\label{fig:wavelength}
\end{figure*}

To verify the linear dispersion of the SAW ($v = f\lambda$), two different pairs of IDTs with different electrode spacing are compared, keeping the metallization ratio fixed such that $m = 0.5$. The IDT geometry is as shown in Fig.~\ref{fig:wavelength} (a). The IDT finger widths $w=d=2~\mu$m and $w=d=3~\mu$m correspond to wavelengths $\lambda =8~\mu$m and $\lambda =12~\mu$m, respectively. The measured resonance frequency $f_\textrm{c}$, labeled with black arrows in Fig.~\ref{fig:wavelength} (b,c) shows the expected inverse linear scaling with wavelength, providing further confirmation of the SAW origin of the resonance feature.

The transmission resonances in both devices show an appreciable hardening as a function of back gate voltage $V_\textrm{bg}$ (Fig.~\ref{fig:wavelength} (b,c)). The rise in acoustic velocity is associated with induced ferroelectric displacement which breaking the inversion symmetry of the crystal structure and couples to the strain field $S$. This phenomenon can be modeled using a Landau-Ginsburg-Devonshire (LGD) free-energy expression, expanding in powers of the displacement $D$ up to the second order (Eq.~\ref{eq:5}) \cite{Alzuaga2014,doi:10.1063/1.2822203,doi:10.1063/1.2837616,doi:10.1063/1.2830866}. For STO when it is paraelectric, the dielectric electro-mechanical response is described within LGD theory \cite{RN4024}.
\begin{eqnarray}
    \label{eq:5}
        F - F_0 =  -pSD + \frac{1}{2}\chi^{-1}D^2 + \frac{1}{2}GSD^2 + \cdots
\end{eqnarray}
In Eq.~\ref{eq:5}, $p$ is the piezoelectric tensor, $\chi$ is the dielectric permittivity tensor and $G$ is the electrostrictive tensor. 

\begin{figure*}
\includegraphics{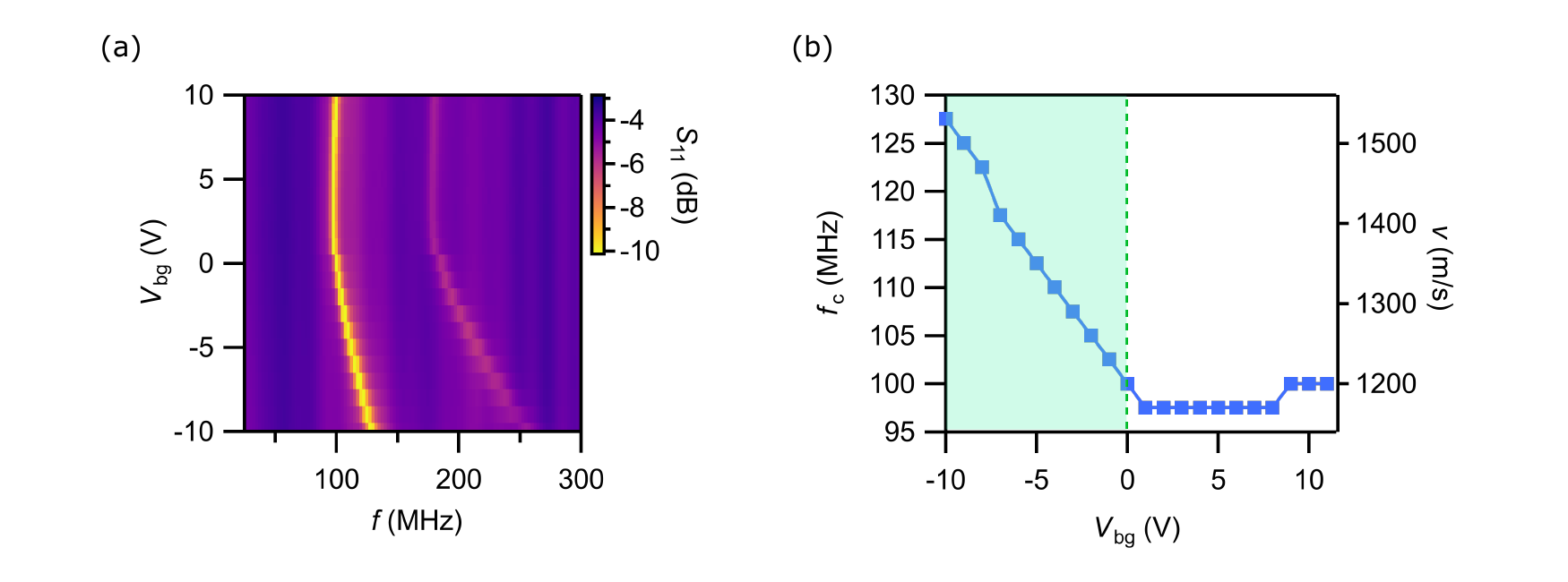}
\caption{Positive and negative $V_\textrm{bg}$ dependence of reflection spectrum and resonance frequency. (a) Reflection spectrum as a function of $V_\textrm{bg}$. There is a resonance dip and a second harmonic dip observed in the spectrum. (b) Resonance center frequency $f_\textrm{c}$ as a function of $V_\textrm{bg}$. The green region highlights where the resonance frequency can be tuned with negative $V_\textrm{bg}$.}
\label{fig:backgate}
\end{figure*}

Surprisingly, the dependence of $f_\textrm{c}$ on $V_\textrm{bg}$ is much stronger when $V_\textrm{bg} < 0$ compared to $V_\textrm{bg} > 0$ (Fig.~\ref{fig:backgate}). This dependence contrasts with pure electrical tuning of the dielectric constant through $V_\textrm{bg}$, in which the tuning effect is symmetric across zero bias \cite{PhysRevB.95.214513}. The LAO thickness is below the critical thickness for spontaneous formation of a conductive LAO/STO interface \cite{Thiel2006-ds}. The interface remains insulating during the experiment, even for the maximum backgate voltage that has been applied, and thus carrier screening or other effects associated with a gate-induced insulator-to-metal transition can be ruled out. One is left with explanations that are tied to the formation and evolution of ferroelastic domains. The motion of such domains under back gate bias is consistent with prior imaging from Honig et al. \cite{Honig2013-uh} which showed that under large negative backgate voltage, tetragonal ferroelastic domains are observed, leading to the anomalously large piezoelectricity at low temperature. The formation and drifting of the ferroelastic domain under negative backgate voltages play an important role coupling with the SAW.

\begin{figure*}
\includegraphics{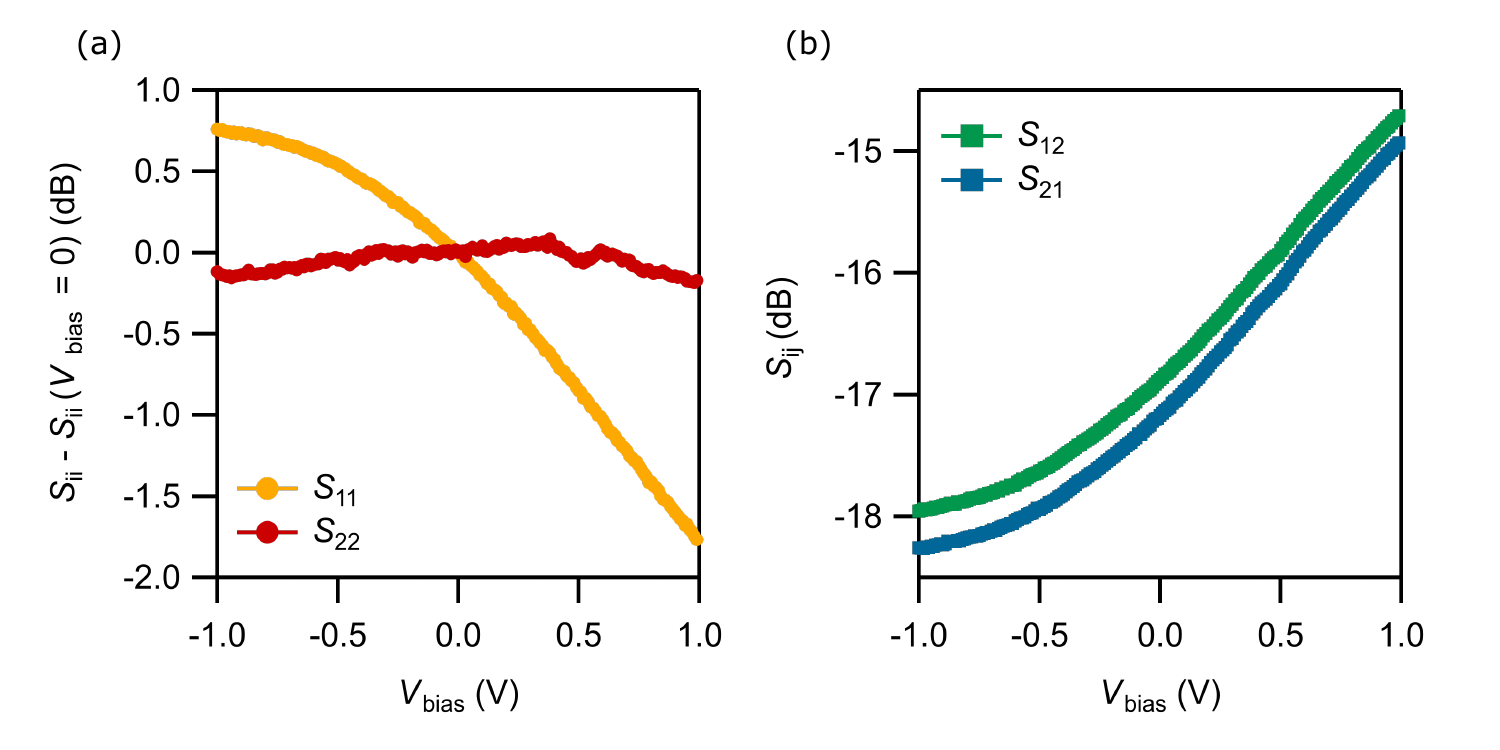}
\caption{Reflection scattering parameters ($S_{11}$, $S_{22}$) measured as a function of $V_\textrm{bias}$ applied across the IDT fingers in port 1, with zero bias at port 2. (a) Reflection amplitude for ports 1 ($S_{11}$) and 2 ($S_{22}$), referenced against the zero-bias value $S_\textrm{ii}~(V_\textrm{bias} = 0)$. (b)Transmission amplitudes ($S_{12}$, $S_{21}$), measured as a function of $V_\textrm{bias}$.}
\label{fig:vbias}
\end{figure*}

To investigate how the magnitude of the SAW coupling depends on the static bias across the IDT, we incorporate a bias tee between the VNA port and IDT connection to apply a dc bias $V_\textrm{bias}$ between IDT neighboring fingers with opposite polarity, and measure the change in the resonance amplitude as a function of $V_\textrm{bias}$. The result (Fig.~\ref{fig:vbias} (a)) shows a quadratic relationship for $S_{11}$ amplitude. The quadratic dependence can be understood as an electrostriction effect in which electric field couples to strain up to the second order (Eq.~\ref{eq:5}). The $V_\textrm{bias}$ induced first-order electric field breaks the inversion symmetry of STO, yielding a linear coupling. The scaling indicates a built-in STO polarization that can be modulated with an applied bias across the IDT. For comparison, port 2 is not subject to a dc bias, and as a result no tuning of the $S_{22}$ amplitude (Fig.~\ref{fig:vbias} (a)) is observed. When the applied $V_\textrm{bias}$ cancels the built-in polarization, the resonance amplitude is minimized, which happens $V_\textrm{bias}\sim -1~\textrm{V}$. Similar tuning is observed for the transmission signals $S_{12}$ and $S_{21}$, as expected.

\section{\label{sec:level1}Discussion and Conclusion}

With an acoustic speed five orders lower than the speed of light, a relatively short acoustic wavelength, and high degree of surface sensitivity, SAW generation, propagation, and detection can be regarded as useful building blocks for manipulating electronic and lattice degrees of freedom in complex-oxide heterostructures and nanostructures. Specifically, SAW has the potential to contribute to quantum information processing architectures \cite{Aref2016} both in superconducting qubits \cite{RN4031,doi:10.1063/5.0023827,PhysRevX.10.021055} and electron spin-based quantum computing architectures \cite{Shilton_1996,doi:10.1063/1.4788826,PhysRevLett.64.2691}. Coupling the superconducting qubits with SAW can help control and measure quantum states \cite{RN4030}. Furthermore, SAW generates moving potential wells with mesoscopic scale which transport electron charges with spin information propagating at speed of sound in a confined one-dimensional channel, helping to meet architectural challenges of long-range transport of spin information \cite{PhysRevB.62.8410,RN4028,RN4029}. At the same time, SAW manipulation of electronic properties may help provide insight into the nature of correlated electronic phases such as superconductivity.

In conclusion, we demonstrate the direct generation and detection of SAW on LAO/STO at cryogenic temperature using superconducting IDTs. Spurious contributions arising from possible bulk acoustic wave components and electronic resonances from the instrument are carefully ruled out. The SAW shows an ultra-low phase velocity which reveals the coupling to the high permittivity of the STO at low temperatures. The SAW quality factor saturates at quantum paraelectric phase transition temperature corroborating the related Q-factor with dielectric permittivity. This method can thus be used to probe behavior near the quantum phase transition. The behavior is consistent with a linear electro-mechanical coupling that is tightly coupled with the ferroelastic domain evolution.

\medskip
\textbf{Supporting Information} \par
Supporting Information is available as Supplementary information.pdf.

\medskip
\textbf{Acknowledgements} \par
JL acknowledges support from the Vannevar Bush Faculty Fellowship program sponsored by the Basic Research Office of the Assistant Secretary of Defense for Research and Engineering and funded by the Office of Naval Research through grant N00014-15-1-2847. The work at University of Wisconsin-Madison was supported by the National Science Foundation under DMREF Grant No. DMR-1629270, AFOSR FA9550-15-1-0334 and AOARD FA2386-15-1-4046. This research is funded by the Gordon and Betty Moore Foundation’s EPiQS Initiative, grant GBMF9065 to C.B.E., Vannevar Bush Faculty Fellowship (N00014-20-1-2844 (C.B.E.). Transport measurement at the University of Wisconsin–Madison was supported by the US Department of Energy (DOE), Office of Science, Office of Basic Energy Sciences (BES), the Materials Sciences and Engineering (MSE) Division under award number DE-FG02-06ER46327.

\medskip
\textbf{Conflict of Interest} \par
The authors declare no conflict of interest.

\medskip
\textbf{Data Availability Statement} \par
Data generated or analysed during this study are included in this published article (and its supplementary information).

\bibliography{Ref.bib}

\end{document}


\maketitle

\newpage
\section{NbTiN superconductivity}
\begin{figure}[ht!]
\includegraphics{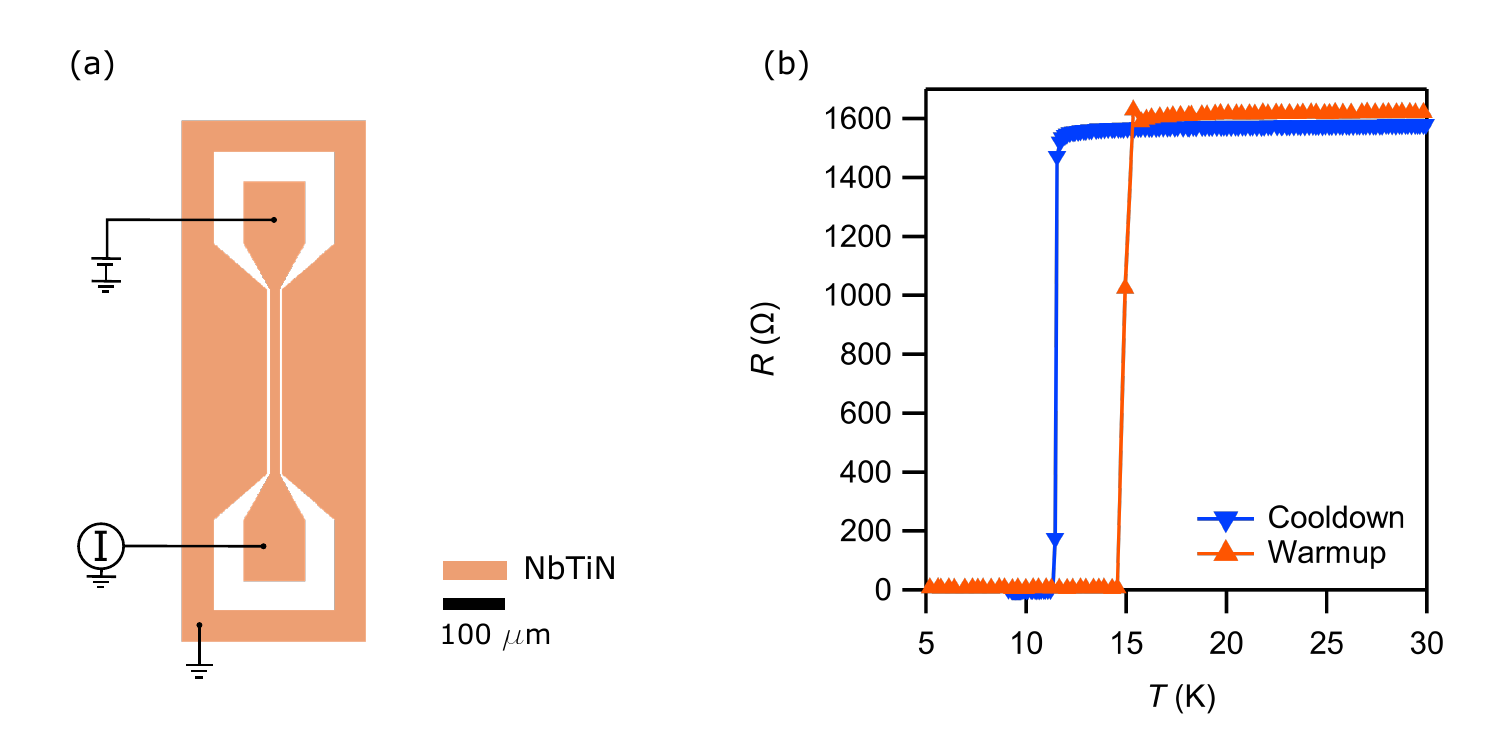}
\centering
\caption{NbTiN superconductivity. (a) Schematic diagram of NbTiN waveguide geometry. (b) Waveguide resistance with respect to the temperature. Blue and orange curves show cooldown and warmup respectively.}
\label{fig:supp1}
\end{figure}

As Fig. \ref{fig:supp1} shows, we study the superconductivity behavior of IDT material NbTiN. We design and pattern a NbTiN waveguide (Fig. \ref{fig:supp1} (a)) and process the transport measurement in a 2 K physical property measurement system (PPMS). Its superconducting critical temperature is in the range of 12 K to 15 K (Fig. \ref{fig:supp1} (b)). The difference can come from the temperature inaccuracy due to relatively fast temperature ramping speed which is 20 K/min here. The resistance changes from normal resistance around 1.6 k$\Omega$ to less than 10 $\Omega$.

